\documentclass[useAMS,usenatbib]{mn2e}
\usepackage{graphicx}
\usepackage{amsmath}
\usepackage{amssymb}

\title{Analytic Expressions for the Surface Brightness Profile of GRB Afterglow Images}

\author[J. Granot]{Jonathan Granot\\
Centre for Astrophysics Research, University of
Hertfordshire, College Lane, Hatfield, Herts, AL10 9AB, UK; j.granot@herts.ac.uk}

\begin{document}

\maketitle

\begin{abstract}

The exact profile of a gamma-ray burst (GRB) afterglow image on the
plane of the sky can provide important constraints on the underlying
physics. In particular, it can test whether the magnetic field in the
emitting shocked external medium decreases significantly with the
distance behind the shock front, or remains roughly
constant. Moreover, it enables more accurate measurements of the
afterglow image size and the expected scintillation properties. In
this work analytic expressions are derived for the afterglow image in
power law segments (PLSs) of the afterglow synchrotron spectrum in
which the emission originates from a very thin layer just behind the
shock, while simple semi-analytic expressions are derived for the
remaining PLSs in which the emission arises from the bulk of the
shocked fluid. In all cases the expressions are for a general power
law external density profile, and are convenient to use in afterglow
studies.

\end{abstract}


\section{Introduction}

The direct measurement of the size of the gamma-ray burst (GRB)
afterglow image in the radio, both through the quenching of
diffractive scintillation
\citep{Frail97} and more directly using the VLBA 
\citep{Taylor04,Taylor05,Pihlstrom07}, provided good support for 
the basic dynamical picture of standard afterglow theory 
\citep{WKF98,ONP04,GR-RL05}. The surface brightness profile 
within the afterglow image can provide additional constraints on the
afterglow physics. It can potentially be tested directly for a
particularly nearby and reasonably bright GRB afterglow, or even for
more distant events in the case of microlensing
\citep[e.g.]{GLS00,GGL01}. In particular, the exact appearance of the
afterglow image on the plane of the sky can be very useful in
improving the accuracy of the afterglow image size measurements, both
when the image is directly but only marginally resolved and its size
is determined through fits to the visibility data, and through the
quenching of scintillation. It can also improve the estimates for the
expected amplitude of scintillation, and thus help in the afterglow
modeling.

The dynamics of GRB afterglows before the jet break time, $t_{\rm
jet}$, are well described by the \citet[][hereafter BM76]{BM76}
self-similar solution, while at later times the exact dynamics of the
GRB jet are much less certain and robust. For this reason, most
calculations of the afterglow image so far
\citep{Waxman97,Sari98,PM98,GPS99a,GPS99b,GL01} have been for the
(quasi-) spherical stage, corresponding to $t < t_{\rm jet}$
\citep[see, however,][]{IN01}. We shall also address the (quasi-) spherical
stage, for the same reasons. In previous works the expressions for the
afterglow image are either not very accurate due to some simplifying
assumption, or not fully analytic, which makes them inconvenient for
others to use. Therefore, in \S~\ref{B_iso} we derive analytic
(\S\S~\ref{sa}, \ref{fc}) or semi-analytic (\S~\ref{bulk}) expressions
for the surface brightness distribution within the afterglow image,
for a general power law external density, and for all of the power law
segments (PLSs) of the afterglow synchrotron spectrum that are
described in \citet[][hereafter GS02]{GS02}, using the notations of
GS02. The magnetic field is assumed to be tangled on small scales with
an isotropic distribution in the comoving frame of the emitting
shocked fluid, and hold a constant fraction of the internal energy
everywhere. The main results are shown in Figures~\ref{fig1} and
\ref{fig2}.

\vspace{-0.5cm}
\section{Deriving Analytic Expressions for the Afterglow Image}
\label{B_iso}

Consider a spherical relativistic blast wave. A photon that is
emitted from the shock front at a lab frame time $t$ when the shock
radius is $R$ and from an angle $\theta$ relative to the line of sight
to the central source (that is located at the origin) reaches the
observer at an observed time
\begin{equation}\label{t_obs}
t_{\rm obs} = t - \frac{R}{c}\cos\theta \ ,
\end{equation}
where $t_{\rm obs} = 0$ corresponds to a photon emitted at the origin
(i.e. the central source) at $t = 0$ [i.e. the time when the outflow
was launched, $R(t=0) = 0$]. For convenience we normalize the shock
radius by its maximal value along the equal arrival time surface of
photons to the observer (along the line of sight, at $\theta = 0$),
$y\equiv R/R_l$. For a power law external density profile, $\rho_{\rm
ext} = Ar^{-k}$ with $k < 4$, the Lorentz factor of the shock front
during the relativistic phase ($\Gamma \gg 1$) scales as $\Gamma
\propto R^{(k-3)/2}$ (BM76), i.e.  $\Gamma =
\Gamma_ly^{(k-3)/2}$. This implies
\begin{equation}\label{R}
t - \frac{R}{c} = \int_0^t(1-\beta)dt \approx
\int_0^R\frac{dR}{2c\Gamma^2} = \frac{R_ly^{4-k}}{2(4-k)\Gamma_l^2c} 
= t_{\rm obs}y^{4-k}\ ,
\end{equation}
and that $R_l = 2(4-k)\Gamma_l^2ct_{\rm obs} = 4(4-k)\gamma_l^2ct_{\rm
obs}$, where $\gamma_l = \Gamma_l/\sqrt{2}$ is the Lorentz factor of
the fluid just behind the shock at $R = R_l$. Since we are interested
in the relativistic regime ($\Gamma \gg 1$), this implies that all the
relevant emission is from small angles ($\theta \ll 1$) so we can
approximate $\cos\theta \approx 1-\theta^2/2$, and equations
\ref{t_obs} and \ref{R} imply 
\begin{equation}
\Gamma_l^2\theta^2 \approx \frac{1-y^{4-k}}{(4-k)y}\ .
\end{equation}
The distance from the center of the image is given by 
\begin{equation}
R_\perp = R\sin\theta \approx
\frac{R_l}{\sqrt{4-k}\,\Gamma_l}\sqrt{y-y^{5-k}}\ ,
\end{equation}
and its maximal value is 
\begin{equation}
R_{\rm\perp,max} = \frac{R_l}{\Gamma_l}(5-k)^{-(5-k)/2(4-k)}\ ,
\end{equation}
and is obtained at $y_* = R_*/R_l = (5-k)^{-1/(4-k)}$ and $\theta_* =
\Gamma_*^{-1} = \Gamma_l^{-1}(5-k)^{-(3-k)/2(4-k)}$, where $\Gamma_* =
\Gamma(R_*)$.  Therefore, the normalized distance from the center of
the circularly symmetric afterglow image is given by
\begin{equation}\label{x}
x \equiv \frac{R_\perp}{R_{\rm\perp,max}} =
\frac{(5-k)^{(5-k)/2(4-k)}}{\sqrt{4-k}}\sqrt{y-y^{5-k}}\ .
\end{equation}

\subsection{Self-Absorbed Spectral Power Law Segments}
\label{sa}

Below the self-absorption frequency $\nu_{\rm sa}$ (PLSs A, B, C in
GS02) the specific intensity (i.e. surface brightness) $I_\nu$ is
equal to the source function ($S_\nu = j_\nu/\alpha_\nu$) at the front
end of the equal arrival time surface, $(5-k)^{-1/(4-k)} \leq y
\leq 1$ ($R_*
\leq R
\leq R_l$), where $y = (5-k)^{-1/(4-k)}$ ($R = R_*$) and $y = 1$ ($R =
R_l$) correspond to $x = 1$ and $x = 0$, respectively:
\begin{equation}\label{I_nu_sa}
I_\nu = \delta^3 I'_{\nu'} = \delta^{3-b}I'_\nu =
\delta^{3-b}\frac{2\nu^2}{c^2}kT_{\rm eff}(\nu)\ ,
\end{equation}
where $I'_{\nu'} \propto (\nu')^b$, $kT_{\rm eff}(\nu)$ is the
effective temperature of the electrons that radiate in the local
rest frame of the emitting fluid just behind the shock at the observed
frequency $\nu$, and
\begin{equation}
\delta \equiv \frac{\nu}{\nu'} 
\approx \frac{2\gamma}{1+\gamma^2\theta^2} \approx 
\frac{2^{3/2}(4-k)\Gamma_l y^{(5-k)/2}}{(7-2k)y^{4-k}+1}\ ,
\end{equation}
is the Doppler factor. Primed and un-primed quantities are measured in
the comoving (emitting fluid) and lab (or observer) frames,
respectively. In PLS A ($\nu_m < \nu < \nu_{\rm sa}$), $kT_{\rm
eff}(\nu) \approx \gamma_e m_e c^2$ for $\gamma_e$ that satisfies $\nu
\approx \nu'_{\rm syn}(\gamma_e) \approx eB'\gamma_e^2/(2\pi m_e c)$,
so that $kT_{\rm eff} \propto (\nu/B')^{1/2} \propto \nu^{1/2}
\Gamma^{-1/2} \rho_{\rm ext}^{-1/4}
\propto \nu^{1/2} y^{3/4}$ [since $(B')^2 \propto e'_{\rm
int} \propto \rho_{\rm ext}\Gamma^2$] and $b = 5/2$. Thus
\begin{eqnarray}
I_\nu &\propto& \nu^{5/2}y^{k/4}\left[1+
\frac{1-y^{4-k}}{2(4-k)y^{4-k}}\right]^{-1/2}
\ ,
\\ \nonumber
\\ \nonumber
\frac{I_\nu(x=1)}{I_\nu(x=0)} &=& \sqrt{\frac{2}{3}}(5-k)^{\frac{-k}{4(4-k)}}
\approx \left\{\begin{matrix} 0.8165 & \ (k=0)\ , 
\cr & \cr 0.6204 & \ (k=2)\ .\end{matrix}\right.
\end{eqnarray}
In PLS B ($\nu < \min[\nu_m,\nu_{\rm sa},\nu_{\rm ac}]$), $kT_{\rm
eff} \approx \gamma_m m_e c^2 \propto \Gamma
\propto y^{(k-3)/2}$, and therefore $b = 2$ and
\begin{eqnarray}
I_\nu &\propto& \nu^2\left[y^{3-k}+
\frac{1-y^{4-k}}{2(4-k)y}\right]^{-1}
\ ,
\\ \nonumber
\\ \nonumber
\frac{I_\nu(x=1)}{I_\nu(x=0)} &=& \frac{2}{3}(5-k)^{\frac{3-k}{4-k}}
\approx \left\{\begin{matrix} 2.2291 & \ (k=0)\ , 
\cr & \cr 1.1547 & \ (k=2)\ .\end{matrix}\right.
\end{eqnarray}

In PLS C ($\nu_{\rm ac} < \nu < \nu_{\rm sa}$) the emission is from
electrons that have suffered significant cooling. Locally, the
electron distribution in this region is approximately mono-energetic,
and the electron Lorentz factor scales as $\gamma_e \propto 1/[(B')^2
l']$ with the distance $l'$ behind the shock front (at the comoving
time of emission) in the comoving frame
\citep{GPS00}. Most of the photons that reach the observer near an
observed frequency $\nu$ originate near $l'_1(\nu,\mu')$ which is
given by $\tau_\nu[l'_1(\nu,\mu')] = 1$, where $\mu' = \cos\theta'$,
and $\theta'$ is the angle between the direction of the photon and the
shock normal (i.e. the radial direction) in the comoving
frame. Therefore, $kT_{\rm eff}(\nu)
\approx \gamma_e[l'_1(\nu,\mu')] m_e c^2$. The path length of a 
photon until it overtakes the shock front (which is receding from the
shocked fluid at a velocity of $\beta'_{\rm sh} c \approx c/3$), $s'$,
is related to its initial distance from the shock front, $l'$, by
\begin{equation}\label{f}
f \equiv \frac{s'}{l'} = \frac{1}{\mu' - \beta'_{\rm sh}} \approx
\frac{3(1+\gamma^2\theta^2)}{2(1-2\gamma^2\theta^2)} \approx
\frac{3}{4}\left[\frac{(7-2k)y^{4-k}+1}{(5-k)y^{4-k}-1}\right]\ ,
\end{equation}
$$
\mu' = \frac{\mu-\beta}{1-\beta\mu} \approx 
\frac{1-\gamma^2\theta^2}{1+\gamma^2\theta^2}\ .
$$
The location of $l'_1$ is where the optically thin and optically thick
fluxes are equal,
\begin{equation}\label{PLS_C}
n's'(l'_1)\frac{P'_{\nu',{\rm max}}}{4\pi}\left(\frac{\nu'}{\nu'_{\rm
syn}[\gamma_e(l'_1)]}\right)^{1/3} =
\frac{2(\nu')^2}{c^2}\gamma_e(l'_1)m_e c^2\ ,
\end{equation}
where $P'_{\nu',{\rm max}} \propto B' \propto R^{-3/2}$, $\nu'_{\rm
syn} \propto B'\gamma_e^2 \propto R^{-3/2}\gamma_e^2$, $n' \propto
\Gamma\rho_{\rm ext} \propto R^{-(3+k)/2}$, and 
$\gamma_e(l'_1) \propto 1/[(B')^2 l'_1]$. In order to use
eq.~(\ref{I_nu_sa}) we evaluate eq.~(\ref{PLS_C}) at $\nu' = \nu$, so
that $l'_1(\nu,\mu') \propto \nu^{5/8}f^{-3/8}(n')^{-3/8}(B')^{-3/2}$
and $kT_{\rm eff}(\nu) \propto \gamma_e[l'_1(\nu)] \propto
f^{3/8}\nu^{-5/8}(n')^{3/8}(B')^{-1/2} \propto
\nu^{-5/8}f^{3/8}y^{(3-3k)/16}$, implying $b = 11/8$ and 
\begin{eqnarray}\nonumber
I_\nu &\propto& \nu^{11/8}y^{(5k-18)/8}\left[1+
\frac{1-y^{4-k}}{2(4-k)y^{4-k}}\right]^{-13/8}
\\
& &\quad\quad\quad\quad\quad\quad\quad\quad
\times\left[\frac{(7-2k)y^{4-k}+1}{(5-k)y^{4-k}-1}\right]^{3/8}\ .
\end{eqnarray}
Note that the surface brightness diverges at the outer edge of the
image ($x = 1$), as $I_\nu \propto (1-x)^{-3/16}$ for $1-x \ll 1$.

\subsection{Fast Cooling Spectral Power Law Segments}
\label{fc}

For PLSs F ($\max[\nu_c,\nu_{\rm sa}] < \nu < \nu_m$) and H ($\nu >
\max[\nu_m,\nu_c,\nu_{\rm sa}]$) the emission comes from electrons
that cool on a time-scale much smaller than the dynamical time, and
therefore it originates from a very thin layer just behind the shock
so we can use the values of $\gamma$, $n'$ and $B'$ just behind the
shock. We have
\begin{equation}
I_\nu = \delta^3 I'_{\nu'} = \delta^{3-b}I'_\nu\ ,\quad
I'_{\nu'} 
\sim s' j'_{\nu'} \sim f\, l'_{c}[\gamma_e(\nu')]\,j'_{\nu'}\ ,
\end{equation}
where $f$ is given by eq.~(\ref{f})\footnote{With the exception that
here the absolute value of the denominator should be used, as is
becomes negative for $y < y_*$ (at the back of the equal arrival time
surface) since the photons (initially) move away from the shock in
this case.}, $l'_c(\gamma_e) \approx 2\pi m_e
^2/[\sigma_T(B')^2\gamma_e]$ is the electron cooling length, and $\nu'
\equiv \nu'_{\rm syn}[\gamma_e(\nu')] \approx eB'\gamma_e^2(\nu')/(2\pi
m_e c)$. Let $L'_{\nu'} \propto R^a(\nu')^b$ be the spectral
luminosity (the total emitted energy of the whole shell [i.e. thin
emitting layer of shock fluid] per unit time and frequency, assuming a
spherical emitting shell), and $P'_{\nu'}$ be the energy emitted per
unit time, volume, and frequency (where both are measured in the
comoving frame). For PLS F, $a = (5-2k)/4$ and $b = -1/2$, while for
PLS H, $a = [14-9p+2k(p-2)]/4$ and $b = -p/2$
\citep[see Table 1 of][]{Granot05}. For isotropic emission in the
comoving frame we have $j'_{\nu'} = P'_{\nu'}/4\pi$ and therefore
$L'_{\nu'} \sim 4\pi R^2 l'_c[\gamma_e(\nu')] P'_{\nu'}
\propto R^2 l'_c[\gamma_e(\nu')] j'_{\nu'} \propto R^2 f^{-1}
I'_{\nu'}$ where both $P'_{\nu'}$ and $j'_{\nu'}$ are evaluated inside
the thin layer of width $l'_c[\gamma_e(\nu')]$ behind the shock front
in which the electrons whose synchrotron frequency is $\nu'$ have not
yet cooled significantly (and $L'_{\nu'}$ is evaluated using the
volume of this layer). Therefore we have $I'_{\nu'} \propto f
R^{a-2}(\nu')^b$ and $I'_\nu \propto f R^{a-2}\nu^b$ so that
\begin{equation}
I_\nu 
\propto \frac{\nu^b\delta^{3-b}f}{y^{2-a}}
\propto \frac{\nu^b\,y^{a+[11-3k-b(5-k)]/2}}
{|(5-k)y^{4-k}-1|\left[(7-2k)y^{4-k}+1\right]^{2-b}}\ .
\end{equation}
In order to express $I_\nu$ as a function of $x$ rather than $y$ we
use eq.~(\ref{x}) to obtain $y(x)$. It is important to note that
$y(x)$ it is double valued, where the two values correspond to the
front ($y_{+} \geq y_*$) and back ($y_{-} \leq y_*$) of the equal
arrival time surface (EATS) of photons to the observer. Here $y_{-}(x)
= R_{-}(x)/R_l$ corresponds to a photon emitted from the back of the
ETAS at a relatively small radius, $R_{-}$, and a large emission
angle, $\theta > 1/\Gamma(R_{-})$, which corresponds to an angle
$\theta_{\rm sh} > 90^\circ$ relative to the radial direction in the
rest frame of the shock front, and therefore initially lags behind the
shock front.  Eventually, at some larger radius $R_{+}$, it catches-up
with the shock front and starts moving ahead of it. From $R_{+}$
onwards its trajectory coincides with that of a photon emitted from
the front of the ETAS [$y_{+}(x) = R_{+}(x)/R_l$] at the exact place
and time when it crossed the shock front, but at an angle $\theta <
1/\Gamma(R_{+})$ that corresponds to $\theta_{\rm sh} > 90^\circ$ so
that it never lags behind the shock front \citep[For more details see
Fig. 1 of][and the related discussion therein]{Granot08}. The two
values, $y_{-}(x)$ and $y_{+}(x)$, coincide at $y_*$ which corresponds
to $x = 1$, i.e. at the outer edge of the image, where the surface
brightness diverges as $I_\nu \propto (1-x)^{-1/2}$ for $1-x \ll 1$
\citep{Sari98,GL01}.  For $k = 3$ the shock Lorentz factor does not
change with radius, the equal arrival time surface becomes an
ellipsoid, and there is a particularly simple solution: $y_\pm =
\frac{1}{2}\left( 1 \pm
\sqrt{1-x^2}\right)$. For the physically interesting case of $k = 2$, 
which corresponds to the stellar wind of a massive star progenitor, we
also obtain an explicit analytic solution: 
\begin{equation}
y_{\pm}(x) = 
\frac{2}{\sqrt{3}}\cos\left[\frac{1}{3}\left(\pi\mp\arctan\sqrt{x^{-4}-1}\right)\right]
\quad({\rm for}\ k = 2)\ .
\end{equation}
The total value of
$I_\nu(x)$ is obtained by summing these two contributions (in
\S~\ref{sa} only the value corresponding to $y_{+}
\geq y_*$ should be used, since the back of the EATS is obscured).

For sufficiently large values $k$ [$k > 32/9 \approx 3.556$ for PLS F,
and $k > 4(9-p)/(10-p)$ for PLS H] the surface brightness diverges at
the center of the image ($x \ll 1$) due to contributions from small
radii ($y \ll 1$), as $I_\nu \propto x^{(32-9k)/2}$ for PLS F and as
$I_\nu \propto x^{[4(9-p)-(10-p)k]/2}$ for PLS H. Physically the
divergence is avoided due to the break down of some underlying
assumption, e.g. for $k > 3$ the shock accelerates and was initially
non-relativistic at some radius $R_{\rm NR}$ corresponding to $y_{\rm
NR} = R_{\rm NR}/R_l$ which introduces a cutoff at $x_{\rm NR} \approx
(y_{\rm NR}/C_k)^{1/2} \ll 1$, where $C_k = (4-k)(5-k)^{-(5-k)/(4-k)}$.

The expression we obtain for $I_\nu(x)$ is slightly different from
that obtained by \citet{Sari98} for $k =0$. The difference arises
since he did not account for the fact that the fluid just behind the
shock moves at a different velocity than the shock front itself. We
can reproduce his results by replacing $f$ in the expression for
$I_\nu$ with $1/|\cos\theta_{\rm sh}|$ where $\theta_{\rm sh}$ is the
angle between the direction to the observer (i.e. that of the emitted
photons that reach the observer) and the shock normal measured in the
rest frame of the shock front (which moves at a velocity of
$\beta'_{\rm sh}c \approx c/3$ relative to the comoving rest frame of
the shocked fluid), i.e.
\begin{equation}
f \to \frac{1}{|\cos\theta_{\rm sh}|} \approx 
\frac{1+\Gamma^2\theta}{|1-\Gamma^2\theta^2|}
\approx \frac{1+2\gamma^2\theta}{|1-2\gamma^2\theta^2|} \approx
\frac{(3-k)y^{4-k}+1}{|(5-k)y^{4-k}-1|}\ .
\end{equation}

\vspace{-0.7cm}
\subsection{Spectral Power Law Segments Originating from 
the Bulk of the Shocked Fluid}
\label{bulk}

In the PLSs that have been treated so far the emission arises from a
very thin layer of shocked fluid just behind the shock front. This has
enabled the use of the values of the hydrodynamic quantities just
behind the shock, and simplified the derivation of the surface
brightness distribution within the afterglow image, $I_\nu(x)$,
resulting in fully analytic expressions for it. The emission in such
PLSs does not depend on the hydrodynamic profile of the shocked fluid
downstream of the shock transition, and responds relatively quickly to
changes in the external density (although the fact that the
contributions to any given observed time are from a wide range of
radii still causes significant smoothing of the observed light curve;
\citealt{NG07}).

Now we turn to calculate $I_\nu(x)$ for PLSs D, G, and E, in which the
emission originates from the bulk of the shocked fluid. In these cases
one must specify the values of the hydrodynamic quantities everywhere
within the region of shocked fluid.  For this purpose we use the
spherical adiabatic self-similar BM76 solution. In this solution the
hydrodynamic variables depend on the self-similar variable $\chi$, and
(GS02)
\begin{equation}\label{x_chi}
x \equiv \frac{R_\perp}{R_{\rm\perp,max}} =
C_k^{-1/2}\sqrt{y - \chi y^{5-k}}\ ,\quad
\chi = \frac{y - C_k x^2}{y^{5-k}}\ ,
\end{equation}
where $C_k = (4-k)(5-k)^{-(5-k)/(4-k)}$, as well as
\begin{eqnarray}
\gamma &=& 2^{-1/2}\Gamma_l y^{(k-3)/2}\chi^{-1/2}\ ,
\\
e' &=& 2\Gamma_l^2\rho_{\rm ext}(R_l)c^2 y^{-3}\chi^{-(17-4k)/[3(4-k)]}\ ,
\\
n' &=& 2^{3/2}\Gamma_l n_{\rm ext}(R_l)y^{-(3+k)/2}\chi^{-(10-3k)/[2(4-k)]}\ ,
\end{eqnarray}
and
\begin{eqnarray}\nonumber
\Gamma_l^2\theta^2 = \frac{1-\chi y^{4-k}}{(4-k)y}\ ,\quad
\gamma^2\theta^2 = \frac{1-\chi y^{4-k}}{2(4-k)\chi y^{4-k}}\ ,
\\
\delta \approx \frac{2\gamma}{1+\gamma^2\theta^2} \approx 
\frac{2^{3/2}(4-k)\Gamma_l\chi^{1/2}y^{(5-k)/2}}{(7-2k)\chi y^{4-k}+1}\ .
\end{eqnarray}
Assuming isotropic emission in the comoving frame, $j'_{\nu'} =
P'_{\nu'}/4\pi$, and using the fact that $I_\nu = \int j_\nu ds$ where
$j_\nu = \delta^2 j'_{\nu'}$ and in our case $ds \approx dR = R_l dy$,
one obtains
\begin{eqnarray}\nonumber
I_\nu(x) = \frac{2(4-k)^2 R_l\Gamma_l^2}{\pi} \int dy
\frac{\chi(y,x) y^{5-k} P'_{\nu'}\left[y,\chi(y,x)\right]}
{\left[(7-2k)\chi(y,x) y^{4-k}+1\right]^2}
\\ \nonumber
\\ \label{I_nu_bulk}
 = \frac{2(4-k)^2 R_l\Gamma_l^2}{\pi}
\int dy\frac{(y-C_k x^2) y^2 P'_{\nu'}(y,x)}{\left[(7-2k)(y-C_k x^2)+y\right]^2}\ .
\end{eqnarray}

Now we need to derive $P'_{\nu'}\left[y,\chi(y,x)\right]$ and
therefore $P'_{\nu'}(y,x)$ for PLSs D, G, and E. For PLSs D ($\nu_{\rm
sa} < \nu < \nu_m < \nu_c$) and G ($\max[\nu_m,\nu_{\rm sa}] < \nu <
\nu_c$), $P'_{\nu'} \sim n' P'_{\nu',{\rm max},e}(\nu'/\nu'_m)^b$ where 
$P'_{\nu',{\rm max},e} \approx \sigma_T m_e c^2 B'/(3e)$ and $\nu'_m =
\nu'_{\rm syn}(\gamma_m) \approx eB'\gamma_m^2/(2\pi m_e c)$ where 
$\gamma_m = g\epsilon_e e'/(n' m_e c^2)$ and $g = (p-2)/(p-1)$ for $p > 2$.
Thus, for PLSs D where $b = 1/3$ and G where $b = (1-p)/2$ we obtain
\begin{eqnarray}\nonumber
P'_{\nu'=\nu/\delta} \propto \nu^b\chi^{[13k-47+b(13+k)]/[6(4-k)]}
y^{[b(4-k)-6-k]/2}
\\
\times\left[(7-2k)\chi y^{4-k} + 1\right]^b\ ,
\end{eqnarray}
\begin{eqnarray}\nonumber
I_\nu \propto \nu^b \int_{y_-(x)}^{y_+(x)} dy
\left(\chi^{[7k-23+b(13+k)]/[6(4-k)]}
\right.\quad\quad\quad\quad\ 
\\ \label{DG}
\quad\quad\quad\quad\ 
\left.\times y^{[b(4-k)+4-3k]/2}
\left[(7-2k)\chi y^{4-k} + 1\right]^{b-2}\right)\ ,
\end{eqnarray}
where the limits of integration over $y$ are the appropriate roots of
the equation $\chi(y,x) = 1$, i.e. $y-y^{5-k} = C_k x^2$ (see
subsection~\ref{fc}).

\begin{figure}
\includegraphics[width=8.5cm]{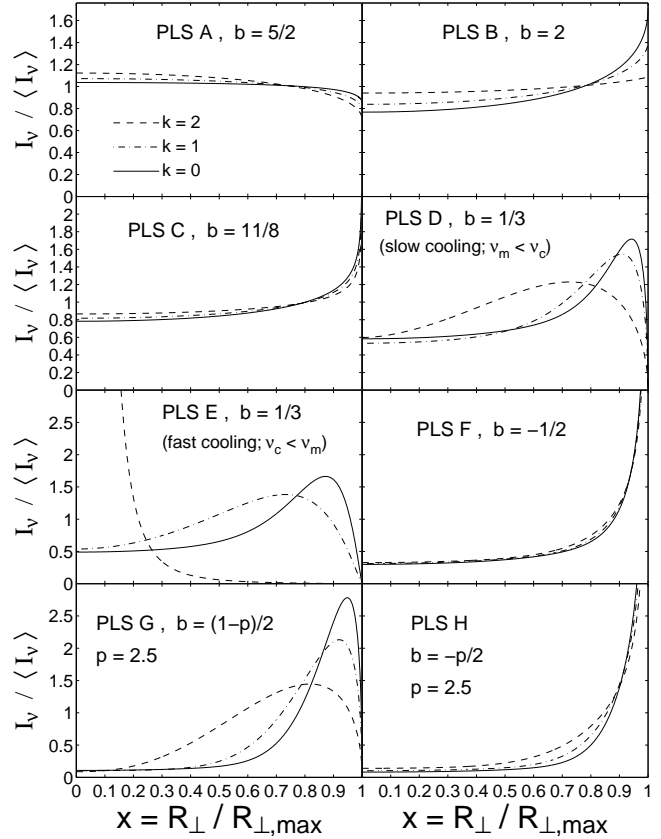}
\caption{The afterglow images: the surface brightness normalized by its 
average value as a function of the normalized distance from the center
of the image, for all of the different power law segments (PLSs) of
the afterglow synchrotron spectrum that are described in GS02,
and for three value of the power law index $k$ of the external
density, where $\rho_{\rm ext} \propto r^{-k}$. For PLS E, the
normalization of the dashed line for $k = 2$ is arbitrary, since the
total flux diverges in that case under our assumptions (see text for
details).}
\label{fig1}
\end{figure}

For PLS E ($\nu_{\rm sa} < \nu < \nu_c < \nu_m$), the emission is
dominated by regions where all of the electrons have cooled
significantly and their energy distribution is practically a delta
function, $N(\gamma_e) \approx n'\delta(\gamma_e-\gamma_{\rm
max}(\chi,y))$, where $\gamma_{\rm max}$ is given by eq.~A12 of
GS02. Using equations A8 and A9 from that paper, one obtains that for
a constant observed time,
\begin{equation}
\gamma_{\rm max}(\chi,y) \propto 
\frac{\chi^{(22-5k)/[6(4-k)]}y^{(1+k)/2}}{\left(\chi^{(19-2k)/[3(4-k)]}-1\right)}\ .
\end{equation}
Therefore, $P'_{\nu'} \approx n' P'_{\nu',{\rm max},e}(\nu'/\nu'_{\rm max})^{1/3}$, where
$\nu'_{\rm max} =
\nu'_{\rm syn}(\gamma_{\rm max}) \approx eB'\gamma_{\rm max}^2/(2\pi m_e c)$.
Altogether, using eq.~(\ref{I_nu_bulk}) we obtain
\begin{eqnarray}\nonumber
I_\nu(x) \propto 
\nu^{1/3}\int dy\left\{y^{(4-5k)/3}
\left[\frac{\chi^{(19-2k)/[3(4-k)]}-1}{\chi^{(18-5k)/[2(4-k)]}}\right]^{2/3}
\right.
\\
\times\,\,
\left.
\left[1+(7-2k)\chi y^{4-k}\right]^{-5/3}\right\}\ .
\end{eqnarray}
It can be shown\footnote{For $x = 0$ we have $\theta = 0$ and $\chi =
y^{k-4}$, and this still approximately holds in the region of interest
here ($\chi \sim y^{k-4} \gg 1$, where $\chi y^{4-k} = 1-C_kx^2/y$
becomes significantly different than 1 only for $y \sim y_{\rm min}
\approx C_kx^2$). Thus $I_\nu \propto \int dy\,y^{2(14-13k)/9}$
becomes dominated by the lower limit of integration, $y_{\rm min}
\approx C_k x^2$, for $k > 37/26$. In this case $I_\nu \propto y_{\rm
min}^{(37-26k)/9} \propto x^{2(37-26k)/9}$.} that for $k > 37/26
\approx 1.423$, $I_\nu(x \ll 1) \propto x^{2(37-26k)/9}$ so that it 
diverges at the center of the image, while the flux in this regime is
dominated by the contribution from $x \ll 1$ and thus scales as $F_\nu
\propto \int_0^1I_\nu(x)xdx \propto 1/(23-13k)$ for $k < 23/13$ and
diverges for $k \geq 23/13 \approx 1.769$. If a lower limit for the
range of integration, $x_{\rm min}$, is introduced then $F_\nu
\propto x_{\rm min}^{4(23-13k)/9}$. Of course, a divergence of 
the total flux is unphysical, and does not really occur. Instead, the
underlying assumptions for this PLS break down in this regime ($k \geq
23/13$), resulting in an introduction of such an $x_{\rm min} \sim
y_{\rm min}^{1/2}$, and PLS E no longer exists in the same form. A
detailed treatment of the interesting behavior in that case is saved
for a separate work.

\begin{figure}
\includegraphics[width=8.5cm]{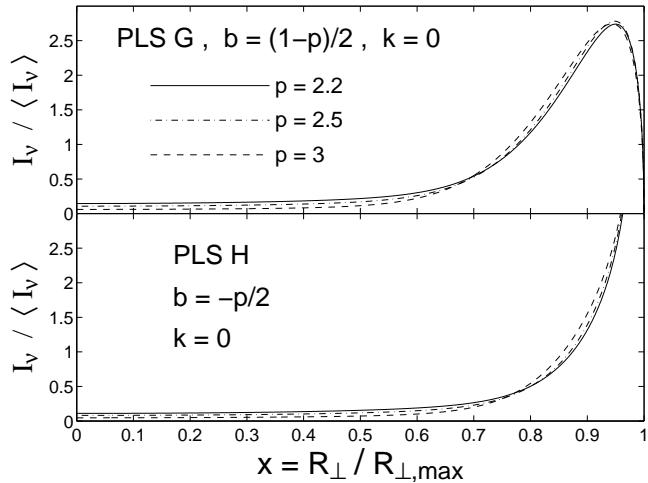}
\caption{This figure demonstrates the dependence of the afterglow 
images on the power-law index, $p$, of the electron energy
distribution, in the two PLSs (G and H) where it has some effect. The
afterglow image becomes somewhat more limb-brightened as the value of
$p$ increases, due to the corresponding decrease in the value of the
spectral index $b$.}
\label{fig2}
\end{figure}

Similarly, in PLS D the surface brightness diverges at the center of
the image for $k > 61/26 \approx 2.346$ as $I_\nu(x \ll 1) \propto
x^{2(61-26k)/9}$, and the flux diverges for $k \geq 35/13
\approx 2.692$. In PLS G the surface brightness diverges 
at the center of the image for\footnote{This is obtained since for
$k(b+1) < -(13b+1)$ the integral in Eq.~(\ref{DG}) becomes dominated
by a narrow range of $y$ values near the lower limit of integration,
where $y \sim y_{\rm min} \approx C_k x^2$ and $\chi \sim (y-y_{\rm
min})/y_{\rm min}^{5-k}$ in the range $(C_k x^2)^{5-k} \lesssim
y-y_{\rm min} \lesssim C_k x^2$, and the lower limit of integration
dominates.} $k > (32-4p)/(11-p)$ as $I_\nu(x \ll 1)
\propto x^{14-5k-(4-k)(p-1)/2}$, and the flux diverges for $k \geq
(36-4p)/(11-p)$. Again, the flux cannot diverge in practice, and this
is an indication that the model assumption break down in those
regimes.

\vspace{-0.3cm}
\section{Discussion}
\label{sec:dis}

Analytic expressions were derived for the afterglow image for PLSs in
which the emission originates from a very thin layer just behind the
shock (\S\S~\ref{sa}, \ref{fc}) while simple semi-analytic expressions
were obtained for the remaining PLSs in which the emission arises from
the bulk of the shocked fluid (\S~\ref{bulk}). These expressions are
for a rather general power law external density profile, $\rho_{\rm
ext} \propto r^{-k}$ with $k < 4$, for which the flow is described by
the BM76 self-similar solution. The relevant expressions are given in
\S~\ref{B_iso}, and illustrated in Figures~\ref{fig1} and \ref{fig2}. 
These expressions fully agree with the afterglow images that were
shown and used in \citet{GL01}, which were calculated using the
formalism of GS02. The flux normalization for the different
PLSs, which also provides the surface brightness normalization, can be
found in Table 1 of GS02.

The magnetic field in the shocked external medium must be considerably
amplified at the shock in order to reproduce the afterglow
observations. However, it is not yet clear how far downstream this
shock produced magnetic field persists. It could in principle decay
considerably at some finite distance behind the shock. If the magnetic
field is significant only within a thin layer (of width $\Delta
\ll R/\gamma^2$) just behind the shock front, and negligible further
downstream from the shock \citep[see, e.g.][]{RR03}, then this will affect
the appearance of the afterglow image. In particular, it will affect
PLSs where the emission would otherwise originate from the bulk of the
shocked fluid (PLSs D, G, and E, which are discussed in
\S~\ref{bulk}). In this case, their appearance would resemble those of 
the fast cooling PLSs, since in both cases the emission arises from a
very thin layer just behind the shock, and the image would become
extremely limb brightened. This is potentially testable in a
microlensing event or if a particularly nearby afterglow image will be
well resolved.

Finally, the afterglow image in general depends also on the magnetic
field structure (its orientation) in the shocked region, not only on
its absolute value (or strength). In this work it was assumed to be
tangled on small scales with an isotropic distribution in the comoving
frame of the emitting shocked fluid. In this case the emission and
absorption coefficients are also isotropic in that frame, which
simplifies the calculation of the afterglow image, and it is very
useful in this respect. However, such a magnetic field structure
predicts no polarization of the afterglow emission, which is
inconsistent with the measurement of linear polarization at the level
of a few percent that has been detected in the optical or NIR
afterglow of several GRBs \citep[see][and references
therein]{Covino04}.  Therefore, in a future work we will discuss the
changes that arise for other possible magnetic field structures in the
emitting region.

\vspace{0.35cm}
The author gratefully acknowledges a Royal Society
Wolfson Research Merit Award.

\end{document}